\documentclass[12pt,a4paper,english,superscriptaddress,aps,nofootinbib]{revtex4}
\usepackage[utf8]{inputenc}
\usepackage[T1]{fontenc}
\usepackage{amsmath,amssymb,graphicx}
\makeatletter
\usepackage{babel}
\usepackage[active]{srcltx}
\usepackage{graphicx,color}
\usepackage{changebar}
\usepackage{hyperref}
\usepackage{color}
\usepackage{xcolor}
\usepackage[T1]{fontenc}
\usepackage{esint}
\usepackage{multirow}
\usepackage{dsfont}
\usepackage{ae}
\usepackage{amsmath}
\usepackage{multirow}
\usepackage{braket}
\usepackage{mathtools}
\usepackage{slashed}
\usepackage{empheq}
\usepackage{multirow}
\usepackage{graphicx}
\usepackage{amsfonts}
\usepackage{amsmath}
\begin{document}
    \title{Codimension-controlled universality of quantum Fisher information singularities at topological band-touching defects}

\author{C. A. S. Almeida}
\email{carlos@fisica.ufc.br}
\affiliation{Departamento de F\'{i}sica, Universidade Federal do Cear\'{a}, 60455-760, Fortaleza, CE, Brazil.}

\begin{abstract}
\vspace{0.25cm}
   \noindent \textbf{Abstract:}
Topological phase transitions in generic multiband systems are mediated
by band-touching defects whose codimension --- the number of momentum
directions along which the gap closes linearly --- varies across
universality classes.
Although singular behavior of fidelity susceptibilities and quantum
Fisher information (QFI) has been computed for specific models, no
unifying principle connecting these results has been identified: it has
remained unclear whether the controlling variable is spatial
dimensionality, band structure, or an intrinsic geometric property of
the defect.
We resolve this question by showing that the singular contribution to
the QFI with respect to the tuning parameter $m$ obeys a universal
power-law scaling $\sim |m|^{p-2}$ for $p \neq 2$, with a logarithmic
divergence $\sim \ln(1/|m|)$ at the marginal codimension $p = 2$,
where $p$ denotes the codimension of the band-touching defect.
This exponent is independent of spatial dimensionality, anisotropies,
ultraviolet regularization, and additional gapped bands, and is
protected by renormalization-group arguments at the linearized fixed
point.
The result unifies previously isolated observations for SSH
chains ($p=1$), Chern insulators ($p=2$), and Weyl semimetals ($p=3$)
as instances of a single codimension-dependent universality class, and
reveals that only defects with $p \leq 2$ generate divergent
information-geometric responses.
This establishes a direct and previously missing link between
topological classification in momentum space and quantum
distinguishability in parameter space, with implications for
metrological sensitivity near topological transitions and for the
experimental detection of topological criticality via quantum geometric
observables.
   
\end{abstract}


\maketitle


\section{Introduction}

Previous computations of fidelity susceptibilities and quantum Fisher
information~(QFI) at topological phase transitions have revealed a
suggestive but unexplained pattern: the strength of information-geometric
singularities decreases systematically as the spatial dimension
increases, from a $1/|m|$ divergence in one-dimensional SSH
chains~\cite{Zanardi2006,Gu2010} to a logarithmic divergence in
two-dimensional Chern insulators~\cite{Yang2007,Albuquerque2010} to a
finite response in three-dimensional Weyl semimetals.
Yet spatial dimensionality and the \emph{codimension} of the
band-touching defect --- the number of momentum directions along which
the gap closes linearly --- are distinct quantities, and no previous
work has identified which of the two governs the universal behavior, or
whether a single geometric quantity controls all cases simultaneously.
As a consequence, these observations have remained as isolated,
model-specific results rather than instances of a unifying principle.
The central difficulty is that fidelity susceptibilities have typically
been computed for specific lattice models in fixed spatial dimensions,
making it impossible to disentangle the role of codimension from other
model-dependent features.
What is needed is a framework that treats the codimension of the
gap-closing manifold as the primary variable, independently of the
embedding spatial dimension, lattice structure, or number of bands.

Topological phase transitions occur through the closing and reopening of energy gaps, where band-touching defects mediate changes in topological invariants such as winding numbers or Chern numbers~\cite{Hasan2010,Qi2011}. In contrast to conventional symmetry-breaking transitions, these critical points are characterized by singular structures in momentum space associated with homotopy classes of maps between parameter manifolds and Bloch-state manifolds~\cite{Nakahara2003,Thouless1982}. Paradigmatic examples include the Su-Schrieffer-Heeger (SSH) chain in one dimension~\cite{Su1979}, Chern insulators in two dimensions~\cite{Haldane1988}, and Weyl semimetals in three dimensions~\cite{Wan2011}.

A complementary geometric perspective is provided by the quantum geometric tensor (QGT), whose real part defines the quantum metric while its imaginary part yields the Berry curvature~\cite{Provost1980,Anandan1990}. The quantum metric has recently attracted considerable attention in condensed matter physics, where it plays a central role in superfluid weight, flat-band superconductivity, and nonlinear transport~\cite{Peotta2015,Ozawa2019}. At the same time, the quantum Fisher information (QFI), which is directly proportional to the quantum metric, quantifies the distinguishability of quantum states under parameter variations and sets the ultimate precision bound in quantum estimation theory~\cite{Braunstein1994,Paris2009}. Singularities of the QFI, corresponding to nonanalytic behavior of the ground-state distinguishability, are well known to signal quantum criticality~\cite{Zanardi2006,Gu2010,You2007,Albuquerque2010}.

For several specific models, fidelity susceptibilities and quantum Fisher information have been computed explicitly. In the SSH chain, the fidelity susceptibility exhibits a $1/|m|$ divergence at the topological transition~\cite{Zanardi2006,Gu2010}. Logarithmic scaling has been reported for two-dimensional Dirac and Chern-type transitions~\cite{Albuquerque2010,Yang2007}, while finite responses arise in higher-dimensional band touchings. These studies, however, focus on particular lattice realizations or spatial dimensionalities. In contrast, the present work identifies the \emph{codimension} of the band-touching defect as the universal quantity controlling the nonanalytic scaling of the QFI, independently of microscopic details or model-specific structures.

In this work, we identify the universal scaling of the nonanalytic part of the QFI at topological band-touching transitions. For generic multiband Hamiltonians linearized near a gap-closing point, we show that the leading singularity of the QFI with respect to the tuning parameter $m$ obeys
\begin{equation}
\mathrm{QFI}_m \sim
\begin{cases}
|m|^{p-2}, & p \neq 2, \\
\ln(1/|m|), & p = 2.
\end{cases}
\label{eq:scaling}
\end{equation}
Remarkably, this exponent depends only on the number of momentum directions in which the gap closes linearly and is independent of anisotropies, ultraviolet regularization, and additional gapped bands. The exponent follows from the universal scaling structure of the linearized critical Hamiltonian and depends only on the codimension of the band-touching defect. The result applies universally to SSH-type transitions ($p=1$), Chern transitions ($p=2$), and Weyl nodes ($p=3$), revealing that only defects of codimension $p \le 2$ generate divergent information-geometric responses.

Our findings demonstrate that topological band-touching defects possess a universal information-geometric fingerprint whose strength is fixed by their codimension. This establishes a direct bridge between topological classification and quantum distinguishability, providing a new geometric lens through which topological phase transitions may be understood.

In this sense, the codimension of a band-touching defect plays a role analogous to spatial dimensionality in conventional critical phenomena, defining an information-geometric universality class for topological phase transitions.

While previous works have analyzed fidelity scaling near quantum critical points, the dependence of the information-geometric exponent on the codimension of topological defects has not been identified as a universal classification principle. In contrast to previous analyses where scaling is tied to spatial dimensionality, here the exponent is controlled by the codimension of the band-touching defect.

\section{General Framework}
\label{sec:framework}

We consider a generic multiband Bloch Hamiltonian $H(\mathbf{k},m)$ depending on momentum $\mathbf{k}$ and on a tuning parameter $m$ controlling a topological phase transition. We assume that the transition occurs through the linear closing of a spectral gap at an isolated point $\mathbf{k}_0$ in the Brillouin zone. Expanding around $\mathbf{k}_0$, we define $\mathbf{q}=\mathbf{k}-\mathbf{k}_0$ and write the low-energy Hamiltonian as
\begin{equation}
H(\mathbf{q},m)
=
\sum_{i=1}^{p} v_i q_i \Gamma_i
+
m \Gamma_{p+1}
+
\delta H(\mathbf{q}),
\label{eq:linearized}
\end{equation}
where $\{\Gamma_a,\Gamma_b\}=2\delta_{ab}$ and $\delta H(\mathbf{q})$ contains higher-order terms in $\mathbf{q}$. The integer $p$ denotes the number of momentum directions along which the gap closes linearly, i.e., the codimension of the band-touching defect.

For an isolated nondegenerate band $|u_0\rangle$, the quantum Fisher information (QFI) with respect to $m$ can be written as
\begin{equation}
g_{mm}(\mathbf{k})
=
\sum_{n\neq 0}
\frac{
|\langle u_n | \partial_m H | u_0 \rangle|^2
}{
(E_n - E_0)^2
}.
\label{eq:qfi_general}
\end{equation}

Near the critical point, only the pair of bands that becomes degenerate contributes nonanalytically to Eq.~(\ref{eq:qfi_general}); all other bands remain separated by a finite gap and yield analytic contributions in $m$. Restricting to the critical subspace and neglecting $\delta H(\mathbf{q})$ for the scaling analysis, the dispersion reads
\begin{equation}
E(\mathbf{q}) = \sqrt{\sum_{i=1}^{p} v_i^2 q_i^2 + m^2}.
\end{equation}

The corresponding metric component takes the universal form
\begin{equation}
g_{mm}(\mathbf{q})
=
\frac{
\sum_{i=1}^{p} v_i^2 q_i^2
}{
4\left(\sum_{i=1}^{p} v_i^2 q_i^2 + m^2\right)^2
}.
\label{eq:metric5}
\end{equation}

The universal structure of Eq. (\ref{eq:metric5}) reflects the geometric scaling of the
quantum metric near a linear band-touching point. Since the low-energy
Hamiltonian depends on momentum only through the combination
$\sqrt{q^2 + m^2}$, the metric density is controlled by a single infrared
scale set by $m$. As a consequence, the singular contribution to the
quantum Fisher information must obey a scaling form 
$QFI_m^{sing} = m^{p-2}\, \Phi(\Lambda/m)$, where $\Phi$ is a dimensionless scaling function determined by the
geometry of the critical Hamiltonian and $\Lambda$ is an ultraviolet
cutoff. The exponent $p-2$ therefore follows from the universal scaling
structure of linear band-touching points and depends only on the
codimension $p$ of the defect.

This scaling structure is analogous to the role played by spatial
dimensionality in conventional critical phenomena, where universal
exponents are determined by the infrared structure of the critical
Hamiltonian.

The total QFI is obtained by integrating the metric density over the Brillouin zone,
\begin{equation}
\mathrm{QFI}_m = \int_{\mathrm{BZ}} d^d k \, g_{mm}(\mathbf{k}),
\end{equation}
and our scaling analysis applies to the nonanalytic (singular) dependence of this integral on $m$.

We define the \emph{singular part} of the QFI as the nonanalytic dependence on $m$ in the limit $m\to 0$. In general, the QFI admits the decomposition
\begin{equation}
\mathrm{QFI}_m
=
A(\Lambda)
+
\mathrm{QFI}_m^{\mathrm{sing}},
\label{eq:decomposition}
\end{equation}
where $A(\Lambda)$ is an analytic background term depending on the ultraviolet cutoff $\Lambda$, and $\mathrm{QFI}_m^{\mathrm{sing}}$ captures the universal infrared contribution.

The singular contribution is governed by
\begin{equation}
\mathrm{QFI}_m^{\mathrm{sing}}
\sim
\int d^p q \,
\frac{q^2}{(q^2+m^2)^2},
\label{eq:qfi_integral}
\end{equation}
up to angular factors. Introducing the scaling transformation $\mathbf{q}=m\mathbf{x}$ gives
\begin{equation}
\mathrm{QFI}_m^{\mathrm{sing}}
\sim
m^{p-2}
\int^{\Lambda/m} d^p x \,
\frac{x^2}{(x^2+1)^2}.
\label{eq:scaled_integral}
\end{equation}

For large $x$, the integrand scales as $x^{p-3}$. Therefore:

\begin{itemize}
\item For $p<2$, the integral is ultraviolet convergent and directly yields
\[
\mathrm{QFI}_m^{\mathrm{sing}} \sim |m|^{p-2}.
\]

\item For $p=2$, the integral produces a logarithmic divergence,
\[
\mathrm{QFI}_m^{\mathrm{sing}} \sim \ln(1/|m|).
\]

\item For $p>2$, the integral diverges in the ultraviolet as $(\Lambda/m)^{p-2}$. After multiplication by $m^{p-2}$, this produces an $m$-independent constant that contributes to the analytic background term $A(\Lambda)$ in Eq.~(\ref{eq:decomposition}). The nonanalytic infrared contribution, arising from the region $x\sim 1$, scales as
\[
\mathrm{QFI}_m^{\mathrm{sing}} \sim |m|^{p-2}.
\]
\end{itemize}

Collecting the results, the singular part of the QFI obeys the universal scaling law
\begin{equation}
\mathrm{QFI}_m^{\mathrm{sing}}
\sim
\begin{cases}
|m|^{p-2}, & p \neq 2, \\
\ln(1/|m|), & p = 2.
\end{cases}
\label{eq:main-result}
\end{equation}
For $p>2$, the total QFI approaches the constant background $A(\Lambda)$ as $m\to0$, with a subleading nonanalytic correction $\sim |m|^{p-2}$. Thus, the critical structure is encoded in the variation with $m$, rather than in the absolute magnitude.

\subsection{Renormalization-group protection of the scaling exponent}

The universal exponent governing the singular behavior of the QFI
is not an artifact of the linearized approximation: it is protected
by the renormalization-group~(RG) structure of the critical fixed
point.
To see this, consider the low-energy effective Hamiltonian near a
band-touching point of codimension~$p$,
\begin{equation}
H(\mathbf{q}, m) = \sum_{i=1}^{p} v_i q_i \Gamma_i + m\Gamma_{p+1}
+ \delta H(\mathbf{q}),
\label{eq:Heff}
\end{equation}
where $\delta H(\mathbf{q})$ collects all corrections beyond linear
order in momentum.
Under a momentum rescaling $\mathbf{q} \to b^{-1}\mathbf{q}$ with
$b > 1$, accompanied by the field rescaling that leaves the kinetic
term invariant, the mass parameter transforms as $m \to b\, m$ while
the leading correction $\delta H \sim q^2$ transforms as
$\delta H \to b^{-1}\delta H$.
Therefore, any perturbation beyond linear order in $\mathbf{q}$ is
\emph{irrelevant} at the linearized fixed point in the RG sense: it
flows to zero under successive rescalings and cannot alter the
infrared behavior of the QFI.

More precisely, the singular part of the QFI is determined by the
infrared structure of the momentum integral
\begin{equation}
\mathrm{QFI}_m^{\mathrm{sing}}
\sim \int \frac{d^p q\; q^2}{(q^2 + m^2)^2},
\label{eq:QFIint}
\end{equation}
which depends on $m$ only through the dimensionless ratio
$q^2/m^2$.
Under the rescaling $\mathbf{q} \to b^{-1}\mathbf{q}$ and
$m \to b\, m$, the integrand is invariant and the measure contributes
a factor $b^{-p}$, so that
\begin{equation}
\mathrm{QFI}_{bm}^{\mathrm{sing}} = b^{-(p-2)}
\,\mathrm{QFI}_m^{\mathrm{sing}},
\label{eq:RGscaling}
\end{equation}
for $p \neq 2$.
Equation~\eqref{eq:RGscaling} is the scaling relation of a
homogeneous function: its unique solution is
$\mathrm{QFI}_m^{\mathrm{sing}} \sim |m|^{p-2}$,
in agreement with the direct calculation of Sec.~\ref{sec:framework}.
For $p = 2$, the exponent vanishes and the scaling relation becomes
$\mathrm{QFI}_{bm}^{\mathrm{sing}} = \mathrm{QFI}_m^{\mathrm{sing}}$,
which is satisfied by a logarithmic dependence
$\mathrm{QFI}_m^{\mathrm{sing}} \sim \ln(1/|m|)$,
the familiar signature of a marginal operator at an upper critical
dimension~\cite{Cardy1996,Sachdev2011}.

Since the only operators that could modify this scaling are irrelevant
at the linearized fixed point, the exponent $p - 2$ is exact within
the universality class defined by linear band touching.
Anisotropies in the velocities $v_i$, additional gapped bands, and
smooth ultraviolet regularizations all correspond to irrelevant
perturbations and therefore do not alter the leading singular behavior.
This places the result on the same theoretical footing as
the universality of critical exponents in conventional
Landau-Ginzburg-Wilson theory~\cite{Wilson1974}, where irrelevant
operators renormalize microscopic parameters but cannot change
the infrared fixed point.
The only perturbations that would modify the exponent are those that
change the codimension~$p$ itself --- for instance, a perturbation
that gaps out one of the $p$ linear directions, reducing the
codimension by one. Such perturbations are \emph{relevant} and
drive the system to a different universality class, consistent with
the picture established here.

\begin{figure}[t]
\centering
\includegraphics[width=\columnwidth]{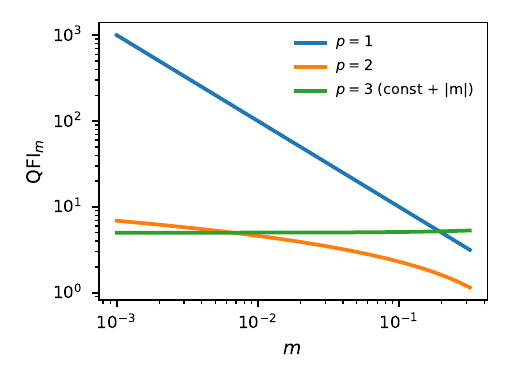}
\caption{
Scaling of the total quantum Fisher information near a topological transition for codimension $p=1,2,3$. For $p=1$ and $p=2$, the QFI is dominated by its singular contribution. For $p=3$, the total QFI approaches a cutoff-dependent constant background with a subleading nonanalytic correction $\sim |m|$.}
\label{fig:scaling}
\end{figure}

Figure \ref{fig:scaling} illustrates the three distinct information-geometric universality classes associated with band-touching defects of codimension $p = 1, 2,$ and $3$. In particular, for $p=3$ the QFI approaches a constant background with a linear correction in $|m|$, while for $p\le2$ the singular contribution dominates.

Importantly, the scaling exponent is insensitive to anisotropies, ultraviolet regularization, and higher-order momentum corrections, being determined solely by the linearized critical structure.



\section{Examples}

We illustrate the universal scaling law in Eq.~(\ref{eq:main-result}) for representative topological transitions in one, two, and three dimensions.

\subsection{One dimension: SSH transition ($p=1$)}

The Su-Schrieffer-Heeger (SSH) model provides the minimal example of a one-dimensional topological transition~\cite{Su1979}. Near the critical point, its Bloch Hamiltonian reduces to
\begin{equation}
H(q,m) = v q \sigma_x + m \sigma_z ,
\end{equation}
where $m$ controls the dimerization and $q$ measures momentum relative to the gap-closing point. The winding number changes as $m$ crosses zero. 

In this case $p=1$, and Eq.~(\ref{eq:qfi_integral}) becomes
\begin{equation}
\mathrm{QFI}_m^{\mathrm{sing}}
\sim
\int dq \,
\frac{q^2}{(q^2+m^2)^2},
\end{equation}
up to angular factors. 

A direct evaluation yields
\begin{equation}
\mathrm{QFI}_m^{\mathrm{sing}}
\sim
\frac{1}{|m|},
\end{equation}
in agreement with Eq.~(\ref{eq:main-result}) for $p=1$. The strong divergence reflects the fact that the gap closes along a single momentum direction, corresponding to a codimension-one defect.

\subsection{Two dimensions: Chern transition ($p=2$)}

A two-dimensional Chern insulator near a topological transition is described by a Dirac Hamiltonian of the form~\cite{Haldane1988}
\begin{equation}
H(\mathbf{q},m)
=
v q_x \sigma_x
+
v q_y \sigma_y
+
m \sigma_z ,
\end{equation}
where the Chern number changes when $m$ changes sign. Here $p=2$, and the singular part of the QFI reads
\begin{equation}
\mathrm{QFI}_m^{\mathrm{sing}}
\sim
\int d^2 q \,
\frac{q^2}{(q^2+m^2)^2}.
\end{equation}
Evaluating the integral yields the logarithmic behavior
\begin{equation}
\mathrm{QFI}_m^{\mathrm{sing}}
\sim
\ln(1/|m|),
\end{equation}
as predicted by Eq.~(\ref{eq:main-result}). The marginal logarithmic divergence reflects the codimension-two nature of the Dirac point mediating the change of Chern number.

\subsection{Three dimensions: Weyl node ($p=3$)}

In three-dimensional Weyl semimetals, a topological transition can occur through the creation or annihilation of Weyl nodes~\cite{Wan2011}. Near a linear band touching, the Hamiltonian reads
\begin{equation}
H(q,m) = v q_x \Gamma_1 + v q_y \Gamma_2 + v q_z \Gamma_3 + m \Gamma_4 ,
\end{equation}
where $m$ controls the separation or annihilation of nodes. In this case $p=3$, and
\begin{equation}
\mathrm{QFI}_m^{\mathrm{sing}}
\sim
\int d^3 q \,
\frac{q^2}{(q^2+m^2)^2}.
\end{equation}

For $p=3$, the momentum integral is finite for any nonzero $m$. 
After ultraviolet regularization, the total QFI approaches a cutoff-dependent constant as $m\to0$, with a subleading nonanalytic correction proportional to $|m|$.

\subsection{Dimensional classification}

These examples confirm that the singular part of the QFI depends solely on the codimension $p$ of the band-touching defect:
\begin{equation}
\mathrm{QFI}_m^{\mathrm{sing}}
\propto
\begin{cases}
|m|^{p-2}, & p \neq 2, \\
\ln(1/|m|), & p = 2.
\end{cases}
\end{equation}
Remarkably, only defects with $p \le 2$ generate divergent information-geometric responses. This establishes a direct correspondence between the dimensional structure of topological band touchings and the strength of quantum distinguishability near the transition.

\section{Discussion and Outlook}

The results established above reveal that topological band-touching defects possess a universal information-geometric signature controlled solely by their codimension $p$. The exponent governing the singular behavior of the quantum Fisher information does not depend on microscopic details, anisotropies, ultraviolet regularization, or the presence of additional gapped bands. Instead, it is fixed by the dimensional structure of the gap-closing manifold.

This finding establishes codimension as an information-geometric universality class. While topological invariants such as winding numbers and Chern numbers classify global properties of Bloch bundles, the quantum Fisher information probes local distinguishability in parameter space. Our analysis shows that these two viewpoints are connected through a universal scaling law: the strength of information-geometric singularities encodes the codimension of the underlying topological defect.

An immediate implication is that only band-touching defects with $p \leq 2$ produce divergent
quantum Fisher information, while higher-codimension defects yield finite responses. In this
sense, information geometry distinguishes between different classes of topological transitions
based on their momentum-space dimensionality. The result also suggests a metrological
interpretation: the detectability of a topological transition through parameter estimation is
fundamentally constrained by the codimension of the critical manifold. This metrological constraint has a transparent physical origin. Near a codimension-one defect, the quantum state varies rapidly with the tuning 
parameter along a single momentum direction, producing a strong and divergent 
sensitivity. As the codimension increases, the critical manifold becomes 
higher-dimensional in momentum space, distributing the state variation over 
more directions and thereby diluting the per-mode contribution to the QFI. 
The threshold at $p = 2$ marks the boundary between divergent and finite 
metrological sensitivity, and may be understood as a dimensional analogue of 
the upper critical dimension in conventional critical phenomena --- but now 
defined in the space of band-touching codimensions rather than in real space.

This perspective also clarifies the role of the logarithmic case $p = 2$. 
The marginal divergence at codimension two is analogous to the logarithmic 
corrections that appear at upper critical dimensions in Landau-Ginzburg 
theory, where mean-field exponents acquire logarithmic modifications. Here 
the logarithm arises from the same mechanism: the integrand is marginally 
non-integrable in the ultraviolet, producing a sensitivity that diverges 
weakly but universally.


Several natural extensions present themselves. First, interacting topological 
systems - where electron-electron interactions renormalize the band structure 
- may shift the location of the critical point but are not expected to alter 
the codimension of the gap-closing manifold, suggesting that the QFI scaling 
law survives in the presence of weak interactions. Second, non-Hermitian band 
touchings, which arise in open quantum systems and exhibit exceptional points 
rather than Dirac-type degeneracies, possess a richer codimension structure 
and may yield new universality classes beyond those identified here. Third, 
Floquet topological phases driven by periodic fields can host band touchings 
at quasienergy zone boundaries whose codimension is controlled by the drive 
parameters, offering a dynamical knob for tuning between universality classes.

From an experimental standpoint, the predicted scaling may be accessed through 
probes sensitive to the quantum geometric tensor. In Dirac and Weyl materials, 
the mass parameter can be tuned via strain engineering, proximity-induced 
magnetic order, or substrate-induced symmetry breaking. The quantum metric is 
known to contribute to the optical conductivity through interband transitions, 
and its singular behavior near a topological transition should be reflected in 
the frequency-integrated optical spectral weight. Similarly, the superfluid 
weight in flat-band superconductors receives a geometric contribution from the 
quantum metric; near a topological transition with $p \leq 2$, this 
contribution should exhibit the divergent scaling identified here. These 
connections suggest that the codimension-dependent universality classes 
predicted by our theory are, in principle, accessible to current experimental 
techniques in quantum materials and cold-atom platforms.

In summary, we have shown that topological band-touching defects exhibit a universal information-geometric fingerprint whose critical exponent is determined exclusively by their codimension. This establishes a direct bridge between topological classification and quantum distinguishability, and provides a compact geometric principle governing the singular behavior of quantum states across topological phase transitions.

\section*{Acknowledgment}

The author would like to express their sincere gratitude to the Conselho Nacional de Desenvolvimento Cient\'{i}fico e Tecnol\'{o}gico (CNPq), and Funda\c{c}\~{a}o Cearense de Apoio ao Desenvolvimento Cient\'{i}fico e Tecnol\'{o}gico (FUNCAP) for their valuable support. He is supported by grants No. 309553/2021-0 (CNPq), 420854/2025-8 (CNPq) and  by Project UNI-00210-00230.01.00/23 (FUNCAP).

\section*{Declaration of Generative AI in Scientific Writing} 
The author used a generative AI tool solely for language refinement and
clarity improvement. All scientific content, derivations, analysis, and conclusions
are entirely the responsibility of the author.

\section*{Conflicts of Interest/Competing Interest}

The author declares that there is no conflict of interest in this manuscript.

\section*{Data Availability Statement}
No Data associated in the manuscript. 

\bibliography{references}

\end{document}